\documentclass{article}

\usepackage{arxiv}

\usepackage[utf8]{inputenc} 
\usepackage[T1]{fontenc}    
\usepackage{hyperref}       
\usepackage{url}            
\usepackage{booktabs}       
\usepackage{amsfonts}       
\usepackage{nicefrac}       
\usepackage{microtype}      
\usepackage{lipsum}
\usepackage{graphicx}
\usepackage{cite}
\graphicspath{ {./images/} }
\usepackage{authblk}

\title{Fast convolutional neural networks for identifying long-lived particles in a high-granularity calorimeter}

\author[1]{Juliette Alimena (juliette.alimena@cern.ch)}
\author[2]{Yutaro Iiyama (yutaro.iiyama@cern.ch)}
\author[3]{Jan Kieseler (jan.kieseler@cern.ch)}
\affil[1]{The Ohio State University, Columbus, OH, USA}
\affil[2]{The University of Tokyo, Tokyo, Japan}
\affil[3]{CERN, Geneva, Switzerland}

\begin{document}
\maketitle
\begin{abstract}
We present a first proof of concept to directly use neural network based pattern recognition to trigger on distinct calorimeter signatures from displaced particles, such as those that arise from the decays of exotic long-lived particles. The study is performed for a high granularity forward calorimeter similar to the planned high granularity calorimeter for the high luminosity upgrade of the CMS detector at the CERN Large Hadron Collider. Without assuming a particular model that predicts long-lived particles, we show that a simple convolutional neural network, that could in principle be deployed on dedicated fast hardware, can efficiently identify showers from displaced particles down to low energies while providing a low trigger rate. 
\end{abstract}

\keywords{Beyond Standard Model \and Hadron-Hadron scattering (experiments) \and Machine learning}

\section{Introduction}


Particles with long lifetimes are an important possibility in the search for new phenomena, and often appear in beyond the standard model theories, notably in models that describe the elementary particle nature of dark matter. When produced at the LHC, these long-lived particles (LLPs) have a distinct experimental signature: they can decay far from the primary proton-proton ($\mathrm{pp}$) interaction but within a detector such as ATLAS or CMS, or even completely pass through the detector before decaying. For example, neutral LLPs could travel a significant distance through the detector before decaying into displaced leptons, photons, or jets~\cite{Aad:2019tcc,Aad:2019kiz,Sirunyan:2019wau,Sirunyan:2019gut,Aaij:2019bvg}.

The data at the ATLAS and CMS experiments are collected using triggers, which select events in real time, reducing the event rate from the 40 MHz bunch crossing rate down to about 1\,kHz that can be written to disk. Most triggers assume that the particles originate from the $\mathrm{pp}$ interaction vertex and are not displaced. Thus, dedicated triggers for displaced particles are necessary to maximize the chances of catching new phenomena at the LHC, in particular for its future data-taking runs. 

The trigger system of the LHC experiments is usually organized in stages. In the CMS experiment, events of interest are selected using a two-level trigger system~\cite{Khachatryan:2016bia}. The first level (L1), composed of custom hardware processors, uses information from the subdetectors and will reduce the data rate to 750\,kHz in CMS at the High-Luminosity LHC (HL-LHC)~\cite{ApollinariG.:2017ojx}, which is planned to start taking data in 2027. In this phase, the upgraded L1 trigger will also feature inputs from the silicon tracker, allowing for real-time track fitting and highly efficient particle-flow reconstruction~\cite{Sirunyan:2017ulk} of objects at the trigger level. The logic will be implemented in field-programmable gate arrays (FPGAs). 

Deep neural networks (DNNs) of limited size can be deployed on FPGAs using dedicated tools such as HLS4ML~\cite{Duarte_2018}, and can therefore now be included directly in the L1 trigger. Given the recent success of DNNs in high energy physics, in particular for complex pattern recognition problems such as b jet identification or heavy flavour jet identification, anomaly detection, as well as shower reconstruction in highly granular calorimeters and particle flow~\cite{guest2018deep,deOliveira:2018lqd,Nguyen:2018ugw,carminati2017calorimetry,komiske2017pileup,CMS-DP-2017-013,CMS-DP-2017-027,ATL-PHYS-PUB-2017-003,TopTaggers,GravNet,JEDINet,ParticleNet,cPFlow,kieseler2020object}, this opens up new possibilities for triggers with simultaneously high computing and physics performance.

Due to the higher occupancy with up to 200 $\mathrm{pp}$ interactions per bunch crossing, in particular in the forward region, a new endcap calorimeter will be installed in CMS for the HL-LHC~\cite{Phase2endcapTDR}. The interleaved HGCal detector layers within the absorber structure will feature a high-granularity electromagnetic section using 28 layers of silicon sensors with pad segmentation, and a hadronic section of 22 layers using the same technology in its innermost layers, and a less segmented scintillator tile section at higher radii. The high granularity of this system will allow for the measurement of particle showers in five parameters: three space dimensions, time, and energy. The HGCal will be the first imaging calorimeter in a running experiment at a high-energy collider, which generates many new opportunities, such as using it for a pattern-recognition-based trigger for displaced particles.

An example of a LLP signature that produces such a displaced, forward signature in form of jets are so-called ``emerging jets''~\cite{Bai:2013xga,Schwaller:2015gea}. Emerging jets contain electrically charged standard model (SM) particles that are consistent with having been created in the decays of new neutral LLPs produced in a parton-shower process by dark quantum chromodynamics (QCD). Dark QCD is a new strong dynamics, similar to SM QCD, but in a separate dark sector. Dark QCD is proposed in order to explain the origin of dark matter~\cite{Bai:2013xga}.

%
This note presents the first proof of concept of using CNN based pattern recognition to trigger on calorimeter signatures, as opposed to an energy over threshold. 
It has been shown in Ref.~\cite{Bhattacherjee_2019} that two dimensional calorimeter images can be used to detect a variety of displaced signatures using convolutional neural networks (CNNs)~\cite{lecun-01a}.
The study presented here is made model independent by investigating the identification of electromagnetic showers that, in general, do not point to the primary $\mathrm{pp}$ interaction vertex. The angle between the projection direction and the particle momentum (angle to projection axis) is in the following referred to as $\alpha$. This trigger improvement will allow us to extend LLP searches with the HL-LHC to low mass and large displacements. 

The study is performed using a toy calorimeter, similar to the HGCal, described in Section~\ref{sec:detector} together with the generated data set. The architecture and training of the DNN is presented in Section~\ref{sec:dnn} and the results are presented in Section~\ref{sec:results}.

\section{Detector and data sample description}
\label{sec:detector}

The endcap calorimeter is built using Geant4~\cite{agostinelli2003geant4} and is placed at $z=3\, \rm m$ distance to the interaction point. It covers a pseudorapidity ($\eta$) between 1.5 and 3.0, has a depth of 34\,cm and consist of 14 equidistant layers. Each layer comprises a 10.4\,mm lead absorber and 300 $\mu\rm m$ silicon sensors. The sensors are placed in 30 rings in $\eta$, each containing 120 segments in $\phi$, leading to 50$\,$400 sensors in total, each with a size of approximately 0.05 in $\eta$ and $\phi$. This configuration corresponds to approximately 60 radiations lengths, and therefore covers electromagnetic showers only. The number of layers and the cell size approximate the granularity of the planned HGCal at first trigger level. Charged particles are subject to a magnetic field of 1~T in $z$ direction.

The signal data set is produced by generating photons at $z=299$\,cm with a flat energy spectrum between 10 and 200\,GeV. The angle with respect to the projection axis is uniformly sampled between 0 and $\pi/3$. The position is randomly set to be within a radius of 20 to 60\,cm with respect to the beam axis. The rotation with respect to the projection axis is also randomly sampled, but constrained such that at least the first and the last layer of the calorimeter are hit. We consider in total 780,000 signal events for training, 8,800 for validation, and 14,400 for evaluating the performance of the proposed algorithm (testing).

To estimate the rate and the effect of multiple interactions per bunch crossing, minimum bias events are produced using Pythia8~\cite{Sj_strand_2008}. We generate two independent samples: 15.3~M events for training and 4~M for testing and validation. The energy deposits of 200 randomly chosen minimum bias events are added to build a background event and to estimate the effect of the contribution of extraneous $\mathrm{pp}$ collisions to the signal. For training and validation, the ratio of signal to background events is 1:1. For testing, 70 background events are generated for each signal event. The rate is calculated by normalising the minimum bias events by the LHC revolution frequency of 11$\,$246 Hz and the number of bunches of 2760~\cite{Evans:2008zzb}.

\section{Neural network and training}
\label{sec:dnn}

\begin{figure}[hbtp]
\begin{center}
\includegraphics[scale=0.7]{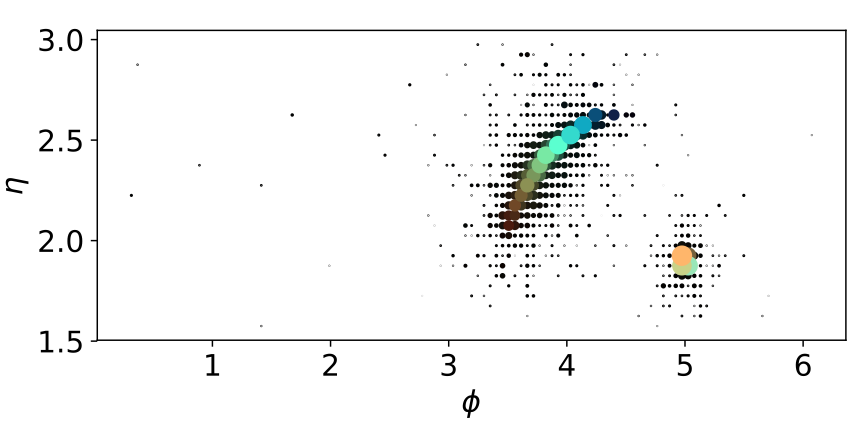}
\caption{A displaced shower (curved line on the left) and a prompt shower (point on the right) in the $\eta$-$\phi$ plane. The calorimeter layers are illustrated using a rainbow color palette, with the color representing the layer number and the marker size indicating the deposited energy.}
\label{fig:llpinpu}
\end{center}
\end{figure}

To distinguish between events with and without a displaced photon, we use a CNN architecture, developed for pattern recognition in images or other data that can be described by a regular grid structure. The detector geometry is unrolled to a 2 dimensional image in $\eta$ and $\phi$ with 14 color dimensions, one for each layer. The first 8 columns from $\phi=0$ to $\phi=0.4$ at $\phi=2*\pi$ are repeated, to account for particles that enter the calorimeter at $\phi \approx 0$. An example of a displaced photon signature after this preprocessing is shown in Figure~\ref{fig:llpinpu}. In this projection, the displaced photon forms a line, while the other particles coming from the primary interaction form points. Moreover, the trajectory of the displaced shower through the layers is distinct from the other particles by a clearly visible color gradient.


The neural network needs to be designed such that it provides a compromise between performance and resource requirements. The latter are particularly stringent if this method should be applied and implemented in dedicated hardware in the first stages of the trigger. While we do not include dedicated studies of the resource requirements on such hardware in this note, the architecture is nevertheless chosen such that it could be adapted to such a setting e.g.~through HLS4ML.

For each pixel, the 14 color dimensions are reduced to 4, by sequentially applying 3 dense neural network layers. The first two layers have 16 nodes, each, and the third has 4. The resulting image embeds the depth information in these 4 features, as opposed to Ref~\cite{Bhattacherjee_2019}, where only a two dimensional representation of the calorimeter deposits is used.
The image containing the encoded depth information is fed through 4 CNN blocks, each containing a CNN layer with a kernel size of $3\otimes 3$ pixels, max pooling and batch normalisation~\cite{ioffe2015batch}. No padding is applied in the neural network. The CNN layer in the first and second block contains 8 filters, and max pooling is applied with a kernel of $2\otimes 2$ pixels. The last two blocks have 12 and 16 filters, and max pooling is only applied on two pixels in $\phi$ direction. The output of the convolutional blocks is flattened and fed through one dense neural network with 32 nodes before the final classifier is calculated using a sigmoid activation. In the other layers, we employ ReLu activations~\cite{relu_cite}. The network contains 10,405 trainable parameters.

The training is performed using tensorflow~\cite{tensorflow} and keras~\cite{keras} within the DeepJetCore framework~\cite{DJC} using the Adam~\cite{kingma2014adam} optimiser. The first epoch is trained with batch size of 50 and a learning rate of 0.0001. The batch size is increased to 500 for another 30 epochs of training with a learning rate of 0.0003.  

\section{Results}
\label{sec:results}

\begin{figure}[htbp]
    \centering
    \includegraphics[width=0.59\textwidth]{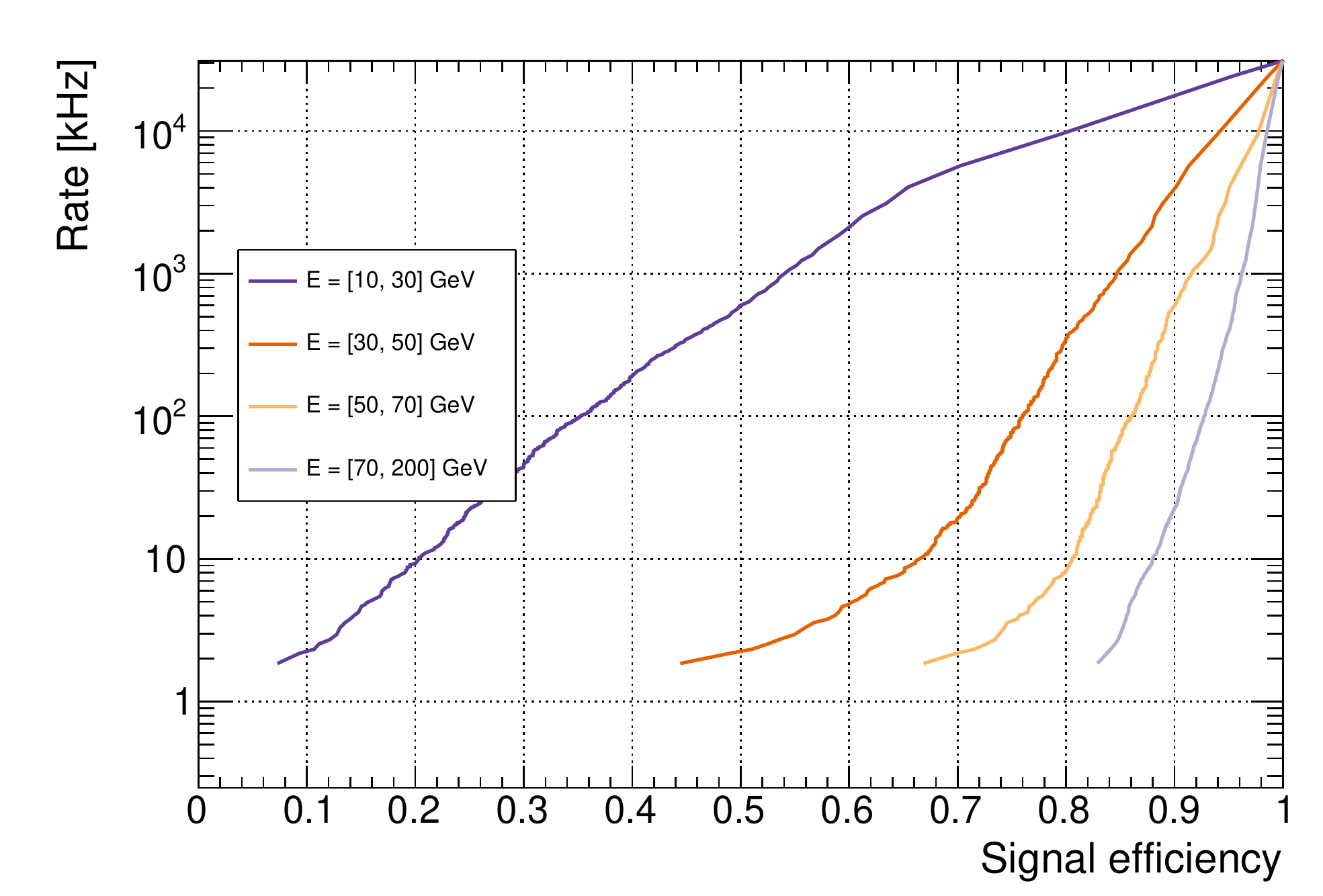}
    \caption{Trigger rate as a function of signal efficiency for different ranges of the photon energy.}
    \label{fig:efficiency}
\end{figure}

We study the efficiency as a function of rate, for different photon energies and angles $\alpha$ with respect to the projection axis. As described in Section~\ref{sec:detector}, both variables are sampled from a uniform distribution. This way of presenting the results is model-independent, whereas any choice of displacement would be inherently model-dependent. As shown in Figure~\ref{fig:efficiency}, the efficiency rapidly increases with the photon energy for a fixed rate, and reaches values above 60\% for a rate of 10\,kHz already for energies larger than 30\,GeV. 

\begin{figure}[hbtp]
    \centering
    \includegraphics[width=0.49\textwidth]{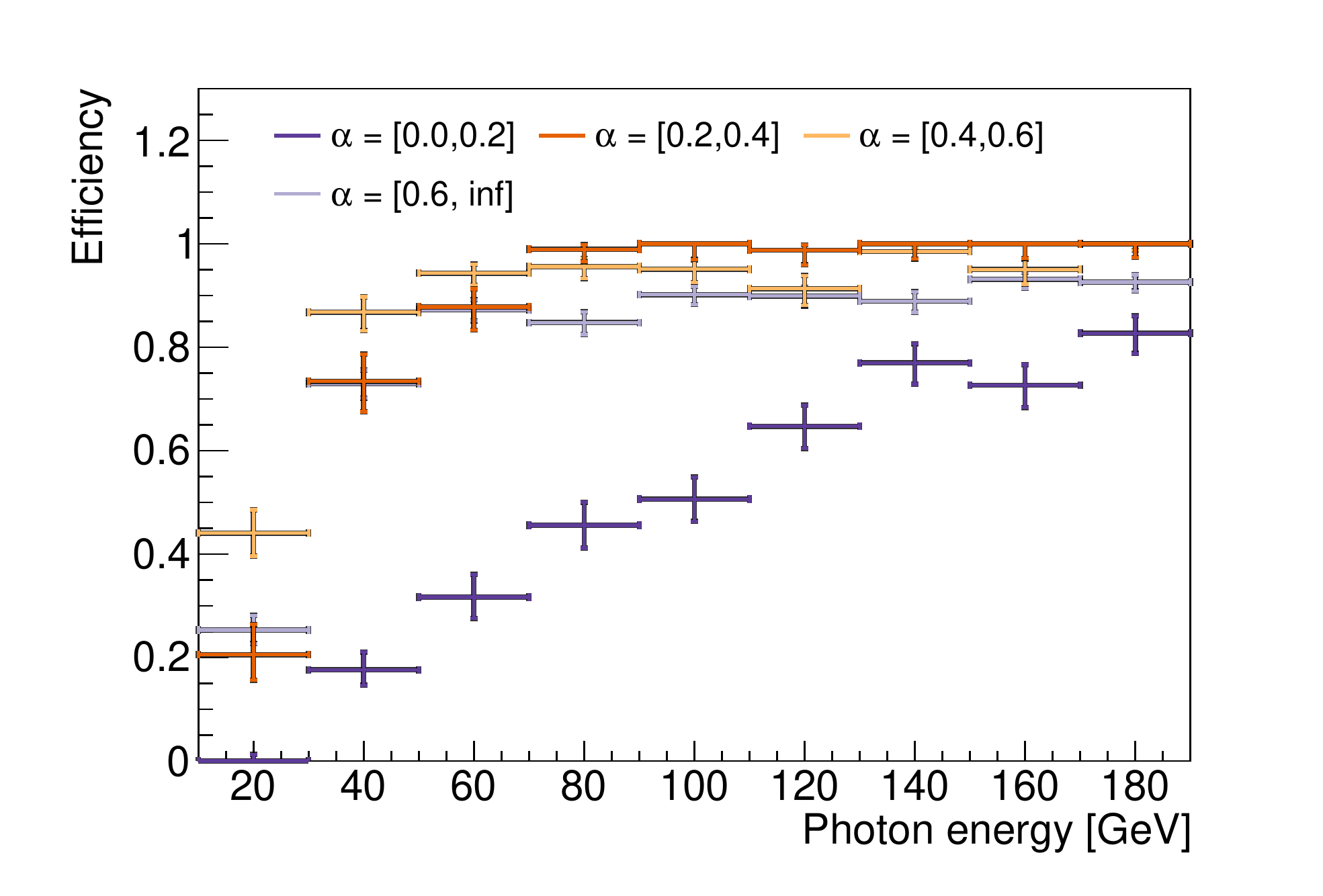}
    \includegraphics[width=0.49\textwidth]{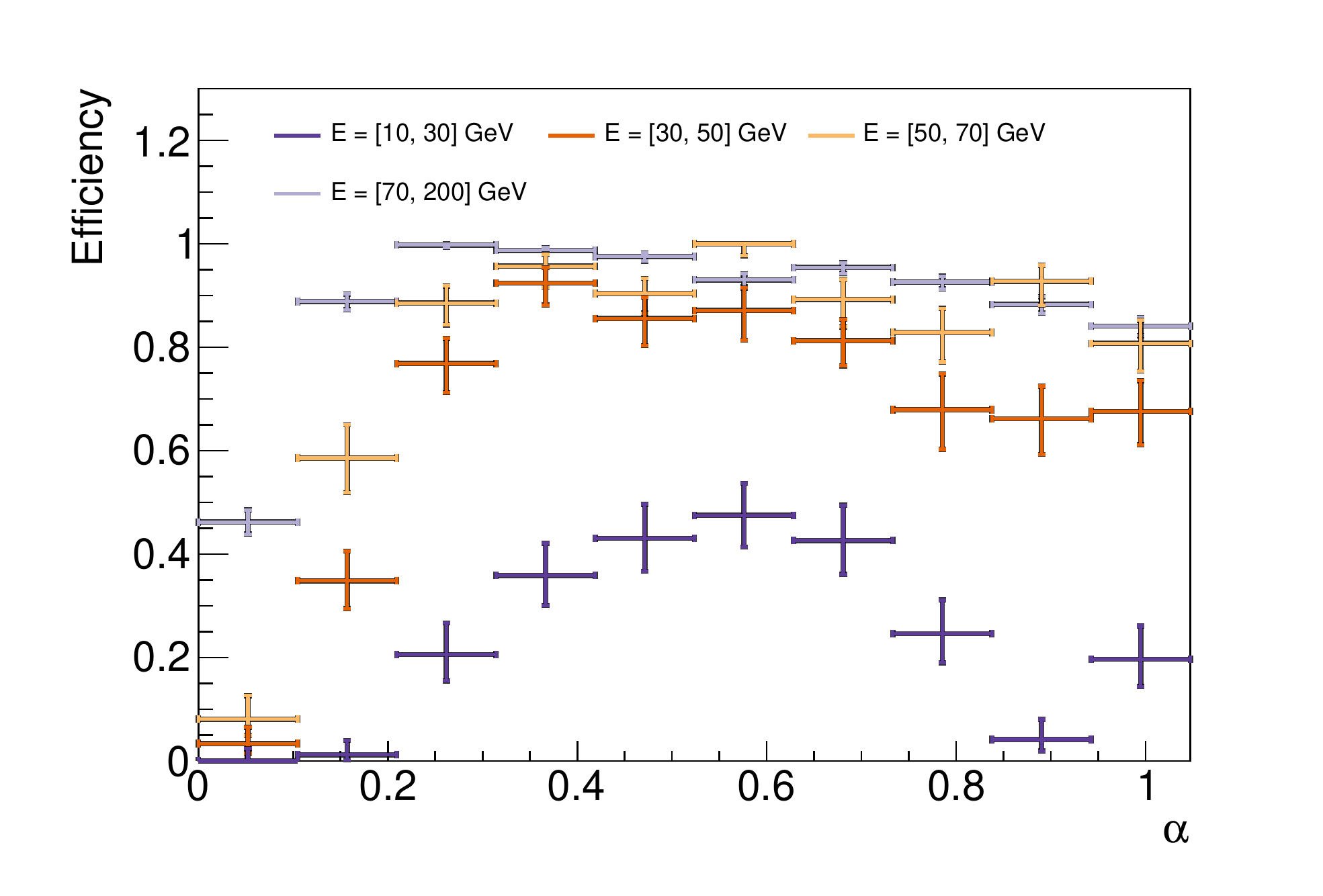}
    \caption{Trigger efficiency for a rate of 15\,kHz; left: as a function of the photon energy for different angles $\alpha$ with respect to the projection axis; right: as a function of $\alpha$ for different photon energies.}
    \label{fig:efficiency_vs}
\end{figure}

As opposed to a trigger that is based on energy thresholds only, the proposed DNN trigger depends critically on the energy and the angle $\alpha$. 
The trigger efficiency as a function of the energy for a trigger rate of 15\,kHz is shown in Figure~\ref{fig:efficiency_vs} left. Particles entering the calorimeter with angles of $\alpha>0.2$ provide a sufficiently distinct signature to be detected already at relatively low energies, while for smaller angles, the efficiency remains moderate up to high energies. 
The dependence of the trigger efficiency on $\alpha$, shown in Figure~\ref{fig:efficiency_vs} right, does not follow the same pattern. Here, the efficiency increases with $\alpha$ for all energies, but decreases slightly beyond approximately $\alpha=0.5$. This behavior is dependent on the DNN architecture and geometry. Starting from a certain angle, the cells hit by a particle are no longer adjacent pixels, but leave a sparse image that can only be resolved by a DNN with sufficient complexity and a larger receptive field.

\section{Summary}

The first proof of concept of using pattern recognition with fast convolutional neural networks to trigger on displaced calorimeter signatures is presented. In particular, displaced signatures in a forward calorimeter can be identified with good efficiency and low false positive rate. For a target trigger rate of 15\,kHz, individual particles with angles with respect to the projection axis greater than $0.2$ can be detected with good efficiency at low particle energy. This study indicates a potential increase in sensitivity to low mass, forward-moving long-lived particles.

\section*{Acknowledgements}

We thank our colleagues in the CERN EP/CMG group for their support using their GPU cluster for training and the preparation of the data sets.


\begin{thebibliography}{10}
\newcommand{\enquote}[1]{``#1''}
\providecommand{\url}[1]{\texttt{#1}}
\providecommand{\urlprefix}{URL }
\expandafter\ifx\csname urlstyle\endcsname\relax
  \providecommand{\doi}[1]{doi:\discretionary{}{}{}#1}\else
  \providecommand{\doi}{doi:\discretionary{}{}{}\begingroup
  \urlstyle{rm}\Url}\fi
\providecommand{\bibinfo}[2]{#2}
\providecommand{\eprint}[2][]{\url{#2}}

\bibitem{Aad:2019tcc}
\bibinfo{author}{{ATLAS Collaboration}}, \enquote{\bibinfo{title}{{Search for
  Displaced Vertices of Oppositely Charged Leptons from Decays of Long-Lived
  Particles in $pp$ Collisions at $\sqrt {s}$ =13 TeV with the ATLAS
  Detector}},} \emph{\bibinfo{journal}{PLB}}, \bibinfo{volume}{801}
  \bibinfo{pages}{135114} (\bibinfo{year}{2020}),
  \doi{10.1016/j.physletb.2019.135114}, \eprint{1907.10037}.

\bibitem{Aad:2019kiz}
\bibinfo{author}{{ATLAS Collaboration}}, \enquote{\bibinfo{title}{{Search for
  Heavy Neutral Leptons in Decays of $W$ Bosons Produced in 13 TeV $pp$
  Collisions Using Prompt and Displaced Signatures with the ATLAS Detector}},}
  \emph{\bibinfo{journal}{JHEP}}, \bibinfo{volume}{10} \bibinfo{pages}{265}
  (\bibinfo{year}{2019}), \doi{10.1007/JHEP10(2019)265}, \eprint{1905.09787}.

\bibitem{Sirunyan:2019wau}
\bibinfo{author}{{CMS Collaboration}}, \enquote{\bibinfo{title}{{Search for
  long-lived particles using delayed photons in proton-proton collisions at
  $\sqrt{s}=$ 13 TeV}},} \emph{\bibinfo{journal}{PRD}}, \bibinfo{volume}{100}
  \bibinfo{pages}{112003} (\bibinfo{year}{2019}),
  \doi{10.1103/PhysRevD.100.112003}, \eprint{1909.06166}.

\bibitem{Sirunyan:2019gut}
\bibinfo{author}{{CMS Collaboration}}, \enquote{\bibinfo{title}{{Search for
  long-lived particles using nonprompt jets and missing transverse momentum
  with proton-proton collisions at $\sqrt{s} =$ 13 TeV}},}
  \emph{\bibinfo{journal}{PLB}}, \bibinfo{volume}{797} \bibinfo{pages}{134876}
  (\bibinfo{year}{2019}), \doi{10.1016/j.physletb.2019.134876},
  \eprint{1906.06441}.

\bibitem{Aaij:2019bvg}
\bibinfo{author}{{LHCb Collaboration}}, \enquote{\bibinfo{title}{{Search for
  $A'\!\to\!\mu^+\mu^-$ Decays}},} \emph{\bibinfo{journal}{PRL}},
  \bibinfo{volume}{124} \bibinfo{pages}{041801} (\bibinfo{year}{2020}),
  \doi{10.1103/PhysRevLett.124.041801}, \eprint{1910.06926}.

\bibitem{Khachatryan:2016bia}
\bibinfo{author}{{CMS Collaboration}}, \enquote{\bibinfo{title}{{The CMS
  Trigger System}},} \emph{\bibinfo{journal}{JINST}}, \bibinfo{volume}{12}
  \bibinfo{pages}{P01020} (\bibinfo{year}{2017}),
  \doi{10.1088/1748-0221/12/01/P01020}, \eprint{1609.02366}.

\bibitem{ApollinariG.:2017ojx}
\bibinfo{author}{G.~Apollinari}, \bibinfo{author}{I.~Béjar~Alonso},
  \bibinfo{author}{O.~Brüning}, \bibinfo{author}{P.~Fessia},
  \bibinfo{author}{M.~Lamont}, \bibinfo{author}{L.~Rossi},
  \bibinfo{author}{L.~Tavian}, \enquote{\bibinfo{title}{{High-Luminosity Large
  Hadron Collider (HL-LHC)}},} \emph{\bibinfo{journal}{CERN Yellow Rep.
  Monogr.}}, \bibinfo{volume}{4} \bibinfo{pages}{1} (\bibinfo{year}{2017}),
  \doi{10.23731/CYRM-2017-004}.

\bibitem{Sirunyan:2017ulk}
\bibinfo{author}{{CMS Collaboration}}, \enquote{\bibinfo{title}{{Particle-flow
  Reconstruction and Global Event Description with the CMS Detector}},}
  \emph{\bibinfo{journal}{JINST}}, \bibinfo{volume}{12} \bibinfo{pages}{P10003}
  (\bibinfo{year}{2017}), \doi{10.1088/1748-0221/12/10/P10003},
  \eprint{1706.04965}.

\bibitem{Duarte_2018}
\bibinfo{author}{J.~Duarte}, \bibinfo{author}{S.~Han},
  \bibinfo{author}{P.~Harris}, \bibinfo{author}{S.~Jindariani},
  \bibinfo{author}{E.~Kreinar}, \bibinfo{author}{B.~Kreis},
  \bibinfo{author}{J.~Ngadiuba}, \bibinfo{author}{M.~Pierini},
  \bibinfo{author}{R.~Rivera}, \bibinfo{author}{N.~Tran},
  \bibinfo{author}{et~al.}, \enquote{\bibinfo{title}{Fast inference of deep
  neural networks in {FPGAs} for particle physics},}
  \emph{\bibinfo{journal}{JINST}}, \bibinfo{volume}{13} \bibinfo{pages}{P07027}
  (\bibinfo{year}{2018}), \doi{10.1088/1748-0221/13/07/p07027},
  \urlprefix\url{http://dx.doi.org/10.1088/1748-0221/13/07/P07027}.

\bibitem{guest2018deep}
\bibinfo{author}{D.~Guest}, \bibinfo{author}{K.~Cranmer},
  \bibinfo{author}{D.~Whiteson}, \enquote{\bibinfo{title}{{Deep Learning and
  its Application to LHC Physics}},} \emph{\bibinfo{journal}{Ann. Rev. Nucl.
  Part. Sci.}}, \bibinfo{volume}{68} (\bibinfo{year}{2018}),
  \doi{10.1146/annurev-nucl-101917-021019}, \eprint{1806.11484}.

\bibitem{deOliveira:2018lqd}
\bibinfo{author}{L.~de~Oliveira}, \bibinfo{author}{B.~Nachman},
  \bibinfo{author}{M.~Paganini}, \enquote{\bibinfo{title}{Electromagnetic
  showers beyond shower shapes},} \emph{\bibinfo{journal}{Nuclear Instruments
  and Methods in Physics Research Section A: Accelerators, Spectrometers,
  Detectors and Associated Equipment}}, \bibinfo{volume}{951}
  \bibinfo{pages}{162879} (\bibinfo{year}{2020}), ISSN
  \bibinfo{issn}{0168-9002}, \doi{10.1016/j.nima.2019.162879},
  \urlprefix\url{http://dx.doi.org/10.1016/j.nima.2019.162879}.

\bibitem{Nguyen:2018ugw}
\bibinfo{author}{T.~Q. Nguyen}, et~al., \enquote{\bibinfo{title}{{Topology
  classification with deep learning to improve real-time event selection at the
  LHC}},}  (\bibinfo{year}{2018}), \bibinfo{note}{{arXiv:1807.00083 [hep-ex]}}.

\bibitem{carminati2017calorimetry}
\bibinfo{author}{D.~Belayneh}, \bibinfo{author}{F.~Carminati},
  \bibinfo{author}{A.~Farbin}, \bibinfo{author}{B.~Hooberman}, et~al.,
  \enquote{\bibinfo{title}{Calorimetry with deep learning: particle
  classification, energy regression, and simulation for high-energy physics},}
  (\bibinfo{year}{2019}), \eprint{1912.06794}.

\bibitem{komiske2017pileup}
\bibinfo{author}{P.~Komiske}, \bibinfo{author}{E.~Metodiev},
  \bibinfo{author}{B.~Nachman}, \bibinfo{author}{M.~Schwartz},
  \enquote{\bibinfo{title}{Pileup Mitigation with Machine Learning (PUMML)},}
  \emph{\bibinfo{journal}{JHEP}}, \bibinfo{volume}{2017}
  (\bibinfo{year}{2017}), ISSN \bibinfo{issn}{1029-8479},
  \doi{10.1007/jhep12(2017)051},
  \urlprefix\url{http://dx.doi.org/10.1007/JHEP12(2017)051}.

\bibitem{CMS-DP-2017-013}
\bibinfo{author}{{CMS Collaboration}}, \enquote{\bibinfo{title}{{CMS Phase 1
  heavy flavour identification performance and developments}},}
  (\bibinfo{year}{2017}), \urlprefix\url{https://cds.cern.ch/record/2263802}.

\bibitem{CMS-DP-2017-027}
\bibinfo{author}{{CMS Collaboration}}, \enquote{\bibinfo{title}{{New
  Developments for Jet Substructure Reconstruction in CMS}},}
  (\bibinfo{year}{2017}), \urlprefix\url{https://cds.cern.ch/record/2275226}.

\bibitem{ATL-PHYS-PUB-2017-003}
\bibinfo{author}{{ATLAS Collaboration}},
  \enquote{\bibinfo{title}{{Identification of Jets Containing $b$-Hadrons with
  Recurrent Neural Networks at the ATLAS Experiment}},}
  (\bibinfo{year}{2017}), \urlprefix\url{https://cds.cern.ch/record/2255226}.

\bibitem{TopTaggers}
\bibinfo{author}{A.~Butter}, \bibinfo{author}{K.~Cranmer},
  \bibinfo{author}{D.~Debnath}, \bibinfo{author}{B.~M. Dillon}, et~al.,
  \enquote{\bibinfo{title}{{The Machine Learning Landscape of Top Taggers}},}
  \emph{\bibinfo{journal}{SciPost Phys.}}, \bibinfo{volume}{7}
  \bibinfo{pages}{014} (\bibinfo{year}{2019}),
  \doi{10.21468/SciPostPhys.7.1.014}, \eprint{1902.09914}.

\bibitem{GravNet}
\bibinfo{author}{S.~Qasim}, \bibinfo{author}{J.~Kieseler},
  \bibinfo{author}{Y.~Iiyama}, \bibinfo{author}{M.~Pierini},
  \enquote{\bibinfo{title}{{Learning representations of irregular
  particle-detector geometry with distance-weighted graph networks}},}
  \emph{\bibinfo{journal}{EPJC}}, \bibinfo{volume}{79} \bibinfo{pages}{608}
  (\bibinfo{year}{2019}), \doi{10.1140/epjc/s10052-019-7113-9},
  \eprint{1902.07987}.

\bibitem{JEDINet}
\bibinfo{author}{E.~Moreno}, \bibinfo{author}{O.~Cerri},
  \bibinfo{author}{J.~Duarte}, \bibinfo{author}{H.~Newman}, et~al.,
  \enquote{\bibinfo{title}{{JEDI-net: a jet identification algorithm based on
  interaction networks}},} \emph{\bibinfo{journal}{Eur. Phys. J.}},
  \bibinfo{volume}{C80} \bibinfo{pages}{58} (\bibinfo{year}{2020}),
  \doi{10.1140/epjc/s10052-020-7608-4}, \eprint{1908.05318}.

\bibitem{ParticleNet}
\bibinfo{author}{H.~Qu}, \bibinfo{author}{L.~Gouskos},
  \enquote{\bibinfo{title}{{ParticleNet: Jet Tagging via Particle Clouds}},}
  (\bibinfo{year}{2019}), \eprint{1902.08570}.

\bibitem{cPFlow}
\bibinfo{author}{F.~A.~D. Bello}, \bibinfo{author}{S.~Ganguly},
  \bibinfo{author}{E.~Gross}, \bibinfo{author}{M.~Kado},
  \bibinfo{author}{M.~Pitt}, \bibinfo{author}{J.~Shlomi},
  \bibinfo{author}{L.~Santi}, \enquote{\bibinfo{title}{Towards a Computer
  Vision Particle Flow},}  (\bibinfo{year}{2020}), \eprint{2003.08863}.

\bibitem{kieseler2020object}
\bibinfo{author}{J.~Kieseler}, \enquote{\bibinfo{title}{Object condensation:
  one-stage grid-free multi-object reconstruction in physics detectors, graph
  and image data},}  (\bibinfo{year}{2020}), \eprint{2002.03605}.

\bibitem{Phase2endcapTDR}
\bibinfo{author}{{CMS Collaboration}}, \enquote{\bibinfo{title}{{The Phase 2
  Upgrade of the CMS endcap calorimeter}},}
  \emph{\bibinfo{journal}{CERN-LHCC-2017-023, CMS-TDR-019}}
  (\bibinfo{year}{2017}),
  \urlprefix\url{http://cds.cern.ch/record/2293646?ln=en}.

\bibitem{Bai:2013xga}
\bibinfo{author}{Y.~Bai}, \bibinfo{author}{P.~Schwaller},
  \enquote{\bibinfo{title}{{Scale of dark QCD}},} \emph{\bibinfo{journal}{Phys.
  Rev. D}}, \bibinfo{volume}{89} \bibinfo{pages}{063522}
  (\bibinfo{year}{2014}), \doi{10.1103/PhysRevD.89.063522}, \eprint{1306.4676}.

\bibitem{Schwaller:2015gea}
\bibinfo{author}{P.~Schwaller}, \bibinfo{author}{D.~Stolarski},
  \bibinfo{author}{A.~Weiler}, \enquote{\bibinfo{title}{{Emerging Jets}},}
  \emph{\bibinfo{journal}{JHEP}}, \bibinfo{volume}{05} \bibinfo{pages}{059}
  (\bibinfo{year}{2015}), \doi{10.1007/JHEP05(2015)059}, \eprint{1502.05409}.

\bibitem{Bhattacherjee_2019}
\bibinfo{author}{B.~Bhattacherjee}, \bibinfo{author}{S.~Mukherjee},
  \bibinfo{author}{R.~Sengupta}, \enquote{\bibinfo{title}{Study of energy
  deposition patterns in hadron calorimeter for prompt and displaced jets using
  convolutional neural network},} \emph{\bibinfo{journal}{JHEP}},
  \bibinfo{volume}{2019} (\bibinfo{year}{2019}), ISSN
  \bibinfo{issn}{1029-8479}, \doi{10.1007/jhep11(2019)156},
  \urlprefix\url{http://dx.doi.org/10.1007/JHEP11(2019)156}.

\bibitem{lecun-01a}
\bibinfo{author}{Y.~LeCun}, \bibinfo{author}{L.~Bottou},
  \bibinfo{author}{Y.~Bengio}, \bibinfo{author}{P.~Haffner},
  \enquote{\bibinfo{title}{Gradient-Based Learning Applied to Document
  Recognition},} \enquote{\bibinfo{booktitle}{Intelligent Signal Processing},}
  \bibinfo{pages}{306}, \bibinfo{publisher}{IEEE Press} (\bibinfo{year}{2001}).

\bibitem{agostinelli2003geant4}
\bibinfo{author}{S.~Agostinelli}, et~al., \enquote{\bibinfo{title}{{GEANT4: A
  Simulation toolkit}},} \emph{\bibinfo{journal}{Nucl. Instrum. Meth. A}},
  \bibinfo{volume}{506} (\bibinfo{year}{2003}),
  \doi{10.1016/S0168-9002(03)01368-8}.

\bibitem{Sj_strand_2008}
\bibinfo{author}{T.~Sjöstrand}, \bibinfo{author}{S.~Mrenna},
  \bibinfo{author}{P.~Skands}, \enquote{\bibinfo{title}{A brief introduction to
  PYTHIA 8.1},} \emph{\bibinfo{journal}{Computer Physics Communications}},
  \bibinfo{volume}{178} \bibinfo{pages}{852} (\bibinfo{year}{2008}), ISSN
  \bibinfo{issn}{0010-4655}, \doi{10.1016/j.cpc.2008.01.036},
  \urlprefix\url{http://dx.doi.org/10.1016/j.cpc.2008.01.036}.

\bibitem{Evans:2008zzb}
\enquote{\bibinfo{title}{{LHC Machine}},} \emph{\bibinfo{journal}{JINST}},
  \bibinfo{volume}{3} \bibinfo{pages}{S08001} (\bibinfo{year}{2008}),
  \doi{10.1088/1748-0221/3/08/S08001}.

\bibitem{ioffe2015batch}
\bibinfo{author}{S.~Ioffe}, \bibinfo{author}{C.~Szegedy},
  \enquote{\bibinfo{title}{Batch Normalization: Accelerating Deep Network
  Training by Reducing Internal Covariate Shift},}
  \enquote{\bibinfo{booktitle}{Proceedings of the 32nd International Conference
  on Machine Learning, {ICML} 2015, Lille, France, 6-11 July 2015},}
  \bibinfo{pages}{448} (\bibinfo{year}{2015}),
  \urlprefix\url{http://proceedings.mlr.press/v37/ioffe15.html}.

\bibitem{relu_cite}
\bibinfo{author}{V.~Nair}, \bibinfo{author}{G.~E. Hinton},
  \enquote{\bibinfo{title}{Rectified Linear Units Improve Restricted Boltzmann
  Machines},} \enquote{\bibinfo{booktitle}{Proceedings of the 27th
  International Conference on International Conference on Machine Learning},}
  ICML’10, \bibinfo{pages}{807}, \bibinfo{publisher}{Omnipress},
  \bibinfo{address}{Madison, WI, USA} (\bibinfo{year}{2010}), ISBN
  \bibinfo{isbn}{9781605589077}.

\bibitem{tensorflow}
\bibinfo{author}{M.~Abadi}, \bibinfo{author}{A.~Agarwal},
  \bibinfo{author}{P.~Barham}, \bibinfo{author}{E.~Brevdo},
  \bibinfo{author}{Z.~Chen}, \bibinfo{author}{C.~Citro},
  \bibinfo{author}{other}, \enquote{\bibinfo{title}{{TensorFlow}: Large-Scale
  Machine Learning on Heterogeneous Systems},}  (\bibinfo{year}{2015}),
  \bibinfo{note}{software available from tensorflow.org},
  \urlprefix\url{https://www.tensorflow.org/}.

\bibitem{keras}
\bibinfo{author}{F.~Chollet}, et~al., \enquote{\bibinfo{title}{Keras},}
  (\bibinfo{year}{2015}), \urlprefix\url{https://github.com/fchollet/keras}.

\bibitem{DJC}
\bibinfo{author}{J.~Kieseler}, \bibinfo{author}{M.~Stoye},
  \bibinfo{author}{M.~Verzetti}, \bibinfo{author}{P.~Silva},
  \bibinfo{author}{S.~S. Mehta}, \bibinfo{author}{A.~Stakia},
  \bibinfo{author}{Y.~Iiyama}, \bibinfo{author}{E.~Bols},
  \bibinfo{author}{S.~R. Qasim}, \bibinfo{author}{H.~Kirschenmann},
  \bibinfo{author}{et~al.}, \enquote{\bibinfo{title}{DeepJetCore},}
  (\bibinfo{year}{2020}), \doi{10.5281/zenodo.3670882}.

\bibitem{kingma2014adam}
\bibinfo{author}{D.~P. Kingma}, \bibinfo{author}{J.~Ba},
  \enquote{\bibinfo{title}{Adam: {A} Method for Stochastic Optimization},}
  \enquote{\bibinfo{booktitle}{3rd International Conference on Learning
  Representations, {ICLR} 2015, San Diego, CA, USA, May 7-9, 2015, Conference
  Track Proceedings},}  (\bibinfo{year}{2015}),
  \urlprefix\url{http://arxiv.org/abs/1412.6980}.

\end{thebibliography}

\end{document}